\documentclass[preprint,aps,nofootinbib]{revtex4}
\usepackage{graphicx}
\usepackage{amsmath}
\usepackage{amssymb}
\begin{document}

\newcommand{\be}{\begin{equation}}
\newcommand{\ee}{\end{equation}}
\newcommand{\bea}{\begin{eqnarray}}
\newcommand{\eea}{\end{eqnarray}}
\newcommand{\mn}{{\mu\nu}}
\newcommand{\tw}{\theta_{W}}

\vspace*{0.4cm}

\title{The Higgs Mechanism and Loop-induced Decays of \\ a Scalar into Two $Z$ Bosons}

\author{\vspace{0.2cm}Qing-Hong Cao$^{a,b}$, C.~B. Jackson$^{a}$, Wai-Yee Keung$^c$, Ian Low$^{a,d}$, and Jing Shu$^{e}$}
\affiliation{\vspace*{0.3cm}
\mbox{$^a$High Energy Physics Division, Argonne National Laboratory, Argonne, IL 60439, USA}  \\ 
\mbox{$^b$Enrico Fermi Institute, University of Chicago, Chicago, IL 60637, USA}  \\
\mbox{$^c$Department of Physics, University of Illinois, Chicago, IL 60607, USA}  \\
\mbox{$^d$Department of Physics and Astronomy, Northwestern University, Evanston, IL 60208, USA}\\
\mbox{$^e$Institute for the Physics and Mathematics of the Universe,}\\
         \mbox{University of Tokyo, Kashiwa, Chiba 277-8568, Japan} \vspace{0.4cm}}

\begin{abstract}
\vspace*{0.5cm} 
We discuss general on-shell couplings of a scalar with two $Z$ bosons using an operator analysis.
In addition to the operator originated from the Higgs mechanism, two dimension-five operators, one
CP-even and one CP-odd, are generated only at the loop-level. Simple formulas are derived for the differential decay
distributions when the $Z$ pair subsequently decay into four leptons by computing the helicity amplitudes, from
which it is shown the CP-odd operator merely induces a phase shift in the azimuthal angular distribution between
the two decay planes of the $Z$ bosons. We also investigate new physics scenarios giving rise to loop-induced decays of
a scalar into $ZZ$ pair, and argue that the total decay width of such a scalar would be order-of-magnitude smaller than that
of a Higgs boson, should such decays be observed in the early running of the LHC. Therefore, the total decay width alone is a strong indicator of
the Higgs nature, or the lack thereof, of a scalar resonance in $ZZ$ final states. In addition, we study the possibility of using the
azimuthal angular distribution to disentangle effects among all three operators.
\end{abstract}

\maketitle

\section{Introduction}

In the standard model (SM), the electroweak gauge bosons obtain their masses through the Higgs mechanism,
which postulates the existence of a scalar particle whose vacuum expectation value (VEV) breaks
the electroweak $SU(2)_L\times U(1)_Y$ symmetry down to $U(1)_{em}$. If the scalar, the Higgs boson, is a $SU(2)_L$ doublet denoted
by $H=(h^+, h)^T$, then its kinetic term
\be
\label{eq:higgskin}
|D_\mu H|^2 = \left| \left(\partial_\mu -i g \frac{\sigma^a}2 W_\mu^a - i g^\prime \frac12 B_\mu \right) H\right|^2
\ee
contains mass terms for electroweak gauge bosons after the neutral component of the
Higgs doublet gets a VEV, $\langle H \rangle = (0, v)^T/\sqrt{2}$, where $g$ and $g^\prime$ are the gauge couplings
for the $SU(2)_L$ and $U(1)_Y$, respectively, and $\sigma^a$ are the Pauli matrices. Using the mass eigenbasis
\be
W^{\pm}_\mu = \frac1{\sqrt{2}}(W^1_\mu \pm i W^2_\mu), \quad Z_\mu = \frac{g W_\mu^3 - g^\prime B_\mu}{\sqrt{g^2+g^{\prime\, 2}}},
\quad A_\mu  = \frac{g^\prime W_\mu^3 + g B_\mu}{\sqrt{g^2+g^{\prime 2}}},
\ee
one finds from Eq.~(\ref{eq:higgskin}) the following masses
\be
m_W= \frac12 g v, \qquad m_Z = \frac12 \sqrt{g^2+g^{\prime\,2}} \, v, \qquad m_A = 0.
\ee
Furthermore, there are also three-point and four-point couplings from the Higgs kinetic term derived by replacing $m_V \to m_V (1+h/v)$
in the gauge boson mass term:
\be
\label{eq:gaugemass}
 \left(1+\frac{h}{v}\right)^2 m_V^2 V_\mu V^\mu,
 \ee
 where $V = W, Z$. The form of the $hVV$ coupling is completely determined by the electroweak gauge invariance to be
 \be
 \label{eq:smhvv}
 -2i\frac{m_V^2}{v} g_{\mu\nu}.
 \ee
 Therefore, measurements of the three-point vertex in Eq.~(\ref{eq:smhvv}) will be a striking confirmation of the Higgs mechanism.

Experimentally, 
the $hVV$ vertex plays an important role in discovering the Higgs boson at the Large Hadron Collider (LHC). For a Higgs mass above 150 GeV or so,  
the branching ratio is dominated by decays into
$WW$ and $ZZ$~\cite{Djouadi:2005gi}. In particular, $h\to ZZ\to 4\ell$ is
the gold-plated mode for the discovery of 
a moderately heavy ($\agt 180$ GeV) Higgs boson, which is a very clean signature with relatively small backgrounds. The excellent
energy resolution of the reconstructed electrons and muons leads to a clear 4-lepton invariant mass peak, which allows for precise
 measurements of the mass and width of the Higgs boson~\cite{Ball:2007zza}.
 
 Given that so far all data from collider experiments
agree with predictions of the SM quite well, there are very few experimental hints on 
what could (and could not) be seen at the LHC.
 Therefore, if a new scalar resonance is observed in the $WW$ and $ZZ$ final states, it is perhaps prudent to proceed without
 presuming the discovery of a Higgs boson whose VEV gives masses to the $W$ and $Z$  bosons. Only until after one could
 verify the decay indeed occurs through  the three-point coupling in Eq.~(\ref{eq:smhvv}), can one gain some confidence in
  the Higgs mechanism as the origin of electroweak symmetry breaking (EWSB).

In this work we study the physics giving rise to decays of a scalar into two $Z$ bosons, with an emphasis
on probing the Higgs nature of the scalar. Such a final state is interesting in its own right because of the high
degree of symmetry in two identical spin-1 particles. Early studies of such systems resulted in
 the Landau-Yang theorem, which forbids decays of a spin-1 particle into two
photons~\cite{Yang:1950rg}. Recently similar arguments to the Landau-Yang theorem have been extended to 
decays of a massive spin-1  particle, the $Z^\prime$ boson, into two $Z$ bosons~\cite{Keung:2008ve}. There it was
discovered that the azimuthal angle between the two decay planes of the $Z$ is a very useful observable in 
discerning different interactions of the $Z^\prime$ with the $Z$ bosons.

Here we consider the production of a scalar $S$ in the gluon fusion channel, which is the dominant production mechanism of a Higgs boson at the
LHC~\cite{Djouadi:2005gi}, and its subsequent decays into two $Z$ bosons. We do not assume the scalar $S$ plays the role of the Higgs boson in 
the Higgs mechanism.
 In particular, we point out non-Higgs-like couplings are induced only at the loop-level, and  investigate in detail
implications on the underlying new physics.
Since we presume the scalar $S$ and the $Z$ bosons are all produced on-shell, implying $m_S \ge 2m_Z$,
our analysis is different and complimentary to studies on anomalous Higgs couplings in the vector boson 
fusion production, where the vector bosons are off-shell~\cite{Hankele:2006ma,Hagiwara:2009wt}.\footnote{ 
It is worth pointing out that at the LHC the production rate in the
gluon fusion channel is  an order of magnitude larger
than the vector fusion production through out a wide range of Higgs mass.} 
(Measurements of anomalous Higgs couplings at the linear collider have been studied in~\cite{Dutta:2008bh}.) Differential distributions of a scalar decaying into
$ZZ\to 4\ell$ in the general case have been computed in Refs.~\cite{Choi:2002jk, Buszello:2002uu}. However, applying the symmetry argument 
as in Refs.~\cite{Yang:1950rg,Keung:2008ve} would allow
us to simplify the decay distributions dramatically, making it transparent the usefulness of the aforementioned azimuthal angle.
We also argue that the total width of a scalar decaying to $ZZ$ through loop-induced effects should be much smaller
than that of a Higgs-like scalar, if the loop-induced decays should be observed at the LHC in the early running. Therefore measurements on the total width alone is
a smoking gun signal for the Higgs nature of the scalar resonance.

This paper is organized as follows: in the next section we compute the differential distribution of the decay of a scalar into $ZZ\to 4\ell$ using the
helicity amplitudes method, followed by a discussion on the possible new physics giving rise to loop-induced couplings. In Section IV we perform
simulations on the total decay width measurements as well as azimuthal angular distributions between the two decay planes of the $Z$ boson. Then
we conclude in Section V. We also provide two appendices, one on a toy model in which the loop-induced coupling is mediated by the heavy $W^\prime$-boson
loop and the other on the Lorentz-invariant construction of the aforementioned azimuthal angle.

\section{HELICITY AMPLITUDES
for $S \to Z(\lambda_1, k_1) Z(\lambda_2, k_2)\to (\ell_1 \bar{\ell}_1)( \ell_2 \bar{\ell}_2)$}

We use the notation $(\lambda_i, k_i), i=1,2$, to denote the helicity state and momentum of the two $Z$ bosons in the laboratory frame. 
Assuming all three particles are on-shell, the possible helicity states $\Psi^{\lambda_1\lambda_2}$ of the $Z$ pair are determined by
conservation of angular momentum to be $\Psi^{++}, \Psi^{--},$ and  $\Psi^{00}$, from which we see the parity-even combinations
are $\Psi^{++}+\Psi^{--}$ and $\Psi^{00}$ while the parity-odd one is $\Psi^{++}-\Psi^{--}$.
In terms of effective Lagrangian, the three helicity amplitudes are described by the following three operators
\be
\label{eq:effeL}
 {\cal L}_{eff}= \frac12\, m_S\, S \left( c_1 Z^\nu Z_\nu
+ \frac12 \frac{c_2} {m_S^2} Z^{\mu\nu}Z_{\mu\nu} +\frac14 \frac{c_3}{
m_S^2} \epsilon_{\mu\nu\rho\sigma} Z^{\mu\nu}Z^{\rho\sigma} \right) ,
\ee
where  
$ Z_{\mu\nu} =\partial_\mu Z_\nu - \partial_\nu Z_\mu $ is the field strength, and $c_i, i=1,2,3,$ are dimensionless constants.
A fourth operator, $Z_{\mu\nu} (Z^\mu \partial ^\nu S - Z^\nu \partial ^\mu S)$, is related to the $c_1$ and $c_2$ terms upon the 
 equation of motion.
The tensor structure of the decay amplitude of 
$S\to Z_1(k_1^\alpha) + Z_2(k_2^\beta)$ is 
\be
\label{eq:tensor}
 \epsilon_1^\alpha \epsilon_2^\beta{\cal M}_{\alpha\beta}=m_S\ \epsilon_1^\alpha \epsilon_2^\beta \left\{ c_1\, g_{\alpha\beta}
           -\frac{c_2}{m_S^2}
     \left[g_{\alpha\beta} (k_1\cdot k_2) -(k_1)_\alpha (k_2)_\beta \right]
     + \frac{c_3}{m_S^2} \epsilon_{\alpha\beta\gamma\delta} k_1^{\gamma} k_2^{\delta} \right\} ,
\ee
where $ \epsilon_1^\alpha$ and $\epsilon_2^\beta$ are the polarization tensors of $Z_1$ and $Z_2$, respectively. Terms
in Eq.~(\ref{eq:tensor}) proportional to $c_2$ and $c_3$ are the so-called anomalous Higgs couplings.

\begin{figure}
\includegraphics[scale=1, angle=0]{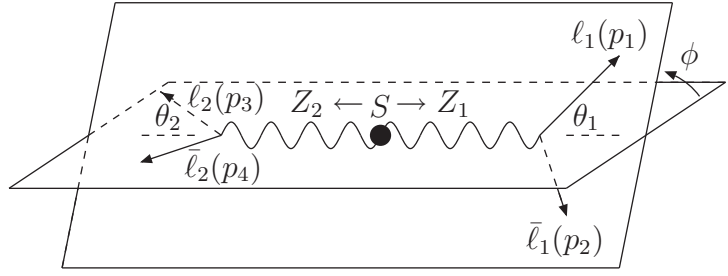}  
\caption{\label{fig1}{\em Two decay planes of $Z_1\to
    \ell_1\bar\ell_1$  and
$Z_2\to \ell_2\bar\ell_2$ define the azimuthal angle $\phi\in[0,2\pi]$
which rotates $\ell_2$ to $\ell_1$ in the transverse view.
The polar angles $\theta_1$ and $\theta_2$ shown are
defined
in the rest frame of $Z_1$ and $Z_2$, respectively.
}}
\end{figure}

Following the method and convention in Ref.~\cite{Keung:2008ve}, 
we calculate the helicity amplitudes ${\cal M}_{\lambda_1 \lambda_2}$:
\bea
 {\cal M}_{\pm\pm} &=& \frac{m_S}2 \left[2 c_1 - c_2 \left(1-\frac{2m_Z^2}{m_S^2}\right)
                         \pm  i c_3 \sqrt{1-\frac{4m_Z^2}{m_S^2}} \ \right] ,\\
  {\cal M}_{00}& =& m_S\left[ c_1 \left(1- \frac{m_S^2}{2m_Z^2}\right)     + c_2 \frac{m_Z^2}{m_S^2}\right] .
  \eea
Notice that 
${\cal M}_{00}$ is real while 
the amplitudes ${\cal M}_{\pm\pm}$ are complex in the presence non-zero $c_3$.  Therefore, we can parametrize the
three helicity amplitudes in terms of two real numbers, ${\cal M}_T$ and ${\cal M}_L$, and one phase $\delta$:
\be
\label{eq:defineM}
 {\cal M}_{++}={M}_T e^{i\delta} \ ,\quad 
    {\cal M}_{--}={M}_T e^{-i\delta} \ ,\quad   {\cal M}_{00}={M}_L  \, ,
\ee    
where
\bea
 { M}_{T} &=& \frac{m_S}2 \left \{ \left[2 c_1 - c_2 \left(1-\frac{2m_Z^2}{m_S^2}\right)  \right]^2
                         + c_3^2 \left(1-\frac{4m_Z^2}{m_S^2}\right) \right \}^{1/2} ,\\
   \delta & =& \arctan \frac{c_3 (1-4m_Z^2/m_S^2)^{1/2}}{2c_1-c_2(1-2m_Z^2/m_S^2)} \ .
\eea
When the two $Z$ bosons further decay into $(\ell_1\bar\ell_1)(\ell_2\bar\ell_2)$, the phase $\delta$ enters into the differential
distribution in a simple way. To see this, recall that the angular distribution of the decay $Z_i \to \ell_i\bar\ell_i$ in the rest frame
of $Z_i$ has the dependence $e^{i m_i \phi_i}$, where $m_i=0,\pm 1$ is the spin projection along the $z$ axis and $\phi_i$ is the
azimuthal angle. Since only the relative angle $\phi$ is physical we set $\phi_1=0$ and $\phi_2=\phi$. (See Fig.~\ref{fig1}.) From Eq.~(\ref{eq:defineM})
we see $\delta$ only enters as a phase shift in $\phi \to \phi +\delta$. Furthermore, the angular dependence of the differential
decay rate is schematically
\be
\label{eq:phidep}
\frac{d\Gamma}{\Gamma d\phi} \sim |a_1 + a_2 e^{i(\phi+\delta)} + a_3 e^{-i(\phi+\delta)}|^2
  \sim b_1 + b_2 \cos (\phi+\delta) + b_3 \cos (2\phi+2\delta) .
\ee
For a $Z^\prime$ boson decaying into the $ZZ$ pair, the $\cos 2\phi$ term is absent in Eq.~(\ref{eq:phidep}) and a similar phase shift $\delta^\prime$
enters as $\phi \to \phi+2\delta^\prime$~\cite{Keung:2008ve}.

Using $g_L$ and $g_R$ to denote the coupling of the $Z$ boson to the left-handed and right-handed leptons, respectively,
we arrive at the differential distribution of $S\to Z_1 Z_2 \to (\ell_1\bar\ell_1)(\ell_2\bar\ell_2)$ following the method of
helicity amplitudes~\cite{Hagiwara:1986vm}:
\bea 
\label{eq:genresult}
 &&{d\Gamma\over \Gamma d\cos\theta_1 d\cos\theta_2 d\phi}
=  \frac1{N}
 \left\{  {1 \over 2}
 \sin^2\theta_1\sin^2\theta_2 \cos(2\phi+2\delta)  
+ \right. \nonumber \\
&&\quad {M_L\over M_T}
     \left[ {1\over2} \sin2\theta_1 \sin2\theta_2 
+ 2\left({g_R^2-g_L^2\over g_R^2+g_L^2}\right)^2
 \sin\theta_1\sin\theta_2 \right]  \cos(\phi+\delta)   +\frac{M_L^2}{M_T^2} \sin^2\theta_1\sin^2\theta_2 \nonumber \\
 && \qquad \left.
     +{1\over2} 
          (1+\cos^2\theta_1)(1+\cos^2\theta_2)
+   2\left({g_R^2-g_L^2\over g_R^2+g_L^2}\right)^2 
             \cos\theta_1\cos\theta_2   \right\} ,
 \eea          
where the definition of $\theta_1, \theta_2,$ and $\phi$ are given in Fig.~\ref{fig1}.
Integrating over the polar angles, we get the expression
\bea 
\label{eq:genphi}
 &&{d\Gamma\over \Gamma d\phi}
= 
 {1\over N}\left\{  \frac89 
  \cos(2\phi+2\delta) 
\right. \nonumber \\ && \qquad \qquad
\left.
 +  \frac{\pi^2}{2} \frac{M_L}{M_T}
      \left({g_R^2-g_L^2\over g_R^2+g_L^2}\right)^2
  \cos(\phi+\delta)  +\frac{16} 9 \left(\frac{M_L^2}{M_T^2}
     +2 \right)
  \right\} .
 \eea     
The normalization factor is given by integrating the above expression,
\be
 N={32\pi \over9 }\left(  \frac{M_L^2}{M_T^2} + 2  \right)  \ .
 \ee
Let's consider turning on $c_i$ one at a time:

\begin{itemize}

\item $c_1\neq 0$ and $c_2=c_3=0$:
\be
\label{eq:c1neq}
\left.{M_L\over M_T}\right|_{c_1\neq 0}
                 =1-\frac{m_S^2}{2m_Z^2}  \qquad {\rm and} \qquad \delta=0 \ .
\ee
This is the case when $S$ plays the role of the Higgs boson in the Higgs mechanism. Since we assume
$m_S \ge 2m_Z$ for on-shell production, we see $|M_L/M_T|_{c_1\neq 0} \ge 1$ and the longitudinal component of the $Z$ dominates
over the transverse components in the decay, especially in the limit of large $m_S$.

\item $c_2\neq 0$ and $c_1=c_3=0$:
\be 
\label{eq:c2neq}
\left.{M_L\over M_T}\right|_{c_2\neq 0} =\frac{-1}{1-m_S^2/(2m_Z^2)} 
                 =-\left(\left.{M_L\over M_T}\right|_{c_1\neq 0}\right)^{-1}  \qquad {\rm and} \qquad \delta=0 \ .
\ee
In this case $|M_L/M_T|_{c_2\neq 0} < 1$ and the transverse polarization of the $Z$ dominates in the decay.

\item $c_3\neq 0$ and $c_1=c_2=0$:
\be
\label{eq:c3neq}
\left.{M_L\over M_T}\right|_{c_3\neq 0} = 0 \qquad {\rm and} \qquad \delta = {\pi \over 2} \ .
\ee 
This is a particularly simple case, as the normalized differential distribution in Eq.~(\ref{eq:genphi}) reduces to
\be
{d\Gamma\over \Gamma d\phi} = \frac{1}{2\pi}\left(1-\frac{1}{4} \cos 2\phi\right) \ .
\ee

\end{itemize}

Previous analysis assuming the SM Higgs boson can be found in 
Ref.~\cite{Barger:1993wt}, while
Refs.~\cite{Chang:1993jy,Han:2000mi} addressed the CP violation due to the simultaneous presence of $c_1$ and $c_3$ terms.
Our general and simple result in Eqs.~(\ref{eq:genresult}) and (\ref{eq:genphi}) is in agreement with 
the lengthy expressions in Ref.~\cite{Buszello:2002uu}. Furthermore, our analysis makes it clear that 
the effect of a non-zero $c_3$, which is CP-odd, is to induce a phase shift in the azimuthal angular distribution.

\section{New Physics and Loop-induced Decays of $S$}

Among the three operators in Eq.~(\ref{eq:effeL}), $c_1$ has the form of the three-point coupling in the Higgs mechanism and could
be present at the tree-level,\footnote{$c_1$ could also be generated through dimension-five operators such as $S|D_\mu H|^2$, which is suppressed
by a high mass scale comparing to Eq.~(\ref{eq:smhvv}). We will not consider this possibility further in this work.} while both $c_2$ and $c_3$ are
higher dimensional operators induced only at the loop level~\cite{Arzt:1994gp}. 
If $c_1=0$ at the tree-level, the scalar 
$S$ is not responsible for giving $W$ and $Z$ bosons a mass. We also assume the existence of the following gluonic operators,
\be
\frac{c_{g2}}{4 m_S}\, S G_{\mu\nu}^a G^{a\, \mu\nu} \qquad {\rm and} \qquad \frac{c_{g3}}{8m_S}\, S \epsilon^{\mu\nu\rho\sigma} G_{\mu\nu}^a G_{\rho\sigma}^a ,
\ee
so as to allow for the production of $S$ in the gluon fusion channel. In the SM it is well-known that $c_{g2}$ is induced by the top-quark triangle loop when $S$ is the Higgs 
boson. In fact, $c_2$ is also present in the SM through the $W$-boson as well as the top-quark loop~\cite{Kniehl:1990mq}, which is nonetheless overwhelmed by
the tree-level $c_1$ given in Eq.~(\ref{eq:smhvv}). On the other hand, the CP-odd operators $c_3$ and $c_{g3}$ can be generated by a fermion triangle loop when the fermion has an axial coupling with the scalar $S$~\cite{Chang:1993jy}.

At the LHC, the event rate $B\sigma(gg\to S \to ZZ)$ is 
\bea
B\sigma(gg\to S \to ZZ)=\sigma (gg \rightarrow S) \times \textrm{Br}(S \rightarrow ZZ) = \sigma (gg \rightarrow S) \times \frac{\Gamma(S \rightarrow ZZ)}{\Gamma_{\rm total}}\  ,
\eea
where the total decay width
$\Gamma_{\rm total}$ is given by summing over all decay channels, including possible decays into SM fermion pair $ \bar{f}f$,
\bea
\quad
\Gamma_{\rm total} =  \sum_{V=g, W, Z, \gamma} \Gamma(S \rightarrow VV)  + \sum_f  \Gamma(S \rightarrow \bar{f}f) \  .
\eea 
A few model-independent observations are in order:
\begin{itemize}

\item While the decay channel into fermions may or may not exist, electroweak symmetry ensures  the existence of decay channels into
$WW$ and $\gamma\gamma$ once $S\to ZZ$ is observed. Establishing the production $gg\to S$ also guarantees a decay channel into two gluons. 

\item If the event rate $B\sigma$ is comparable to the SM expectation of a Higgs boson, then 
the branching ratio into $ZZ$ pair should be sizable
\be
\label{eq:brsize}
\textrm{Br}(S \rightarrow ZZ) \agt {\cal O}(10^{-1})\ .
\ee
The SM Higgs production and decay $gg\to h\to ZZ\to 4\ell$ has an event rate in the order of 5 fb after multiplying $\sigma\times {\rm Br}$ with the pre-selection 
efficiency~\cite{Ball:2007zza}.
Therefore if $\sigma \times {\rm Br}$ is an order-of-magnitude smaller that that of a SM Higgs, it would require an integrated luminosity of 300 fb$^{-1}$ to achieve
5$\sigma$ significance for discovery, which is clearly beyond the early running of the LHC.

\item If $S\to ZZ$ is observed to occur through the loop-induced operators in the early LHC data, then a sizable
$\textrm{Br}(S \rightarrow ZZ)$ implies the total decay width
\be
\Gamma_{\rm tot} = \frac{\Gamma(S\to ZZ)}{{\rm Br}(S\to ZZ)} 
\ee 
should also be one-loop suppressed.

\item Similarly, loop-induced $S\to ZZ$ and a sizable
$\textrm{Br}(S \rightarrow ZZ)$ imply\footnote{In Eq.~(\ref{eq:brbound}) we have neglected an extra factor for decaying into massive gauge bosons, which is order unity
unless $m_S$ is very close to the $2m_Z$ threshold.} 
\be
\label{eq:brbound}
 \frac{\Gamma(S \rightarrow ZZ)}{\Gamma(S\to gg)} \sim {\cal O}\left({c_i^2 \over c_{gi}^2}\right) \ \agt \
 {\cal O}(10^{-1})  \ .
\ee
Since we expect $c_i$ and $c_{gi}$ to be proportional to the electroweak coupling $\alpha_{\rm ew}$ and the strong coupling $\alpha_s$, respectively, 
 a large multiplicity factor ($\agt {\cal O}(1)$) in $c_i$ should be present.

\end{itemize}
In the following we investigate new physics scenarios where the production and decay into $ZZ$ of $S$ occur predominantly through loop-induced 
operators. Such possibilities arise naturally if $S$ is a SM singlet and couples to SM matter only through a messenger sector.
In particular we focus on cases with a sizable branching ratio ${\rm Br}(S\to ZZ)$ as in Eq.~(\ref{eq:brsize}), so that $S$ would have a comparable event
rate to that of a SM Higgs boson.

\subsection{Fermion Loop-induced $S\to gg$}

In the SM gluon fusion production is induced by the top quark loop~\cite{Djouadi:2005gi},
\be
c_{g2}^{(SM)}=\frac{\sqrt{2}\alpha_s}{3\pi} \frac{m_S}{v}\ .
\ee
It is well-known that this coefficient is
related to the top contribution to the gluon two-point function from the Higgs low-energy theorem~\cite{Ellis:1975ap}. If the messenger sector contains
 a pair of heavy vector-like fermions $(Q^c, Q)$ in the fundamental representation of $SU(3)_c$ with the interaction
\be
m_Q Q^c Q + y_Q S\, Q^c Q,
\ee
then its contribution to the gluon two-point function is 
\be
\label{eq:s2gg}
-\frac14\left[1- {g_s^2 \over 16\pi^2} b^{(3)}_F \log {{M}_Q^2({\cal S}) \over \mu^2}\right] G_{\mu\nu}^a G^{a\,\mu\nu},
\ee
where $b^{(3)}_F= 2/3$ is the contribution to the one-loop beta function of QCD from a Dirac fermion and ${M}_Q({\cal S})=m_Q + y_Q {\cal S}$ is the mass of the new heavy fermion $Q$ when turning on the scalar as a background field $S\to {\cal S}$.
To obtain the scalar-gluon-gluon coupling, the Higgs low-energy theorem instructs us to expand Eq.~(\ref{eq:s2gg}) to the first order in ${\cal S}$~\cite{Low:2009di}:
\be
c_{g2} = {\alpha_s \over 3\pi} {m_S \over m_Q} y_Q.
\ee
Strictly speaking, the low-energy theorem applies only when the mass of the particle in the loop is much larger than the scalar mass,
$m_S^2/(4m_Q^2)\ll 1$, so that
the loop diagram can be approximated by a dimension-five operator. We will always work in this limit in the present study. The partial width
of $S\to gg$ can be computed:
\be
\label{eq:sZZg}
\Gamma(S\to gg) = {1 \over 8\pi}\, c_{g2}^2\, m_S = {\alpha_s^2 \over 72\pi^3}  {m_S^3 \over m_Q^2} y_Q^2.
\ee

\subsection{Fermion Loop-induced $S\to ZZ$}

Next we consider the case when the messenger sector contains vector-like fermions $(L^c, L)$ charged under the electroweak gauge group with the interaction
\be 
m_L L^c L + y_L S\, L^c L,
\ee
where $(L^c, L)$ are in the fundamental representation of $SU(2)_L$ and carry the hypercharge $Y_L$ under $U(1)_Y$. The contribution of 
$L$ to the two-point function of the $Z$ boson is simply
\be
-\frac14\left[1- {e^2 N_c \over 16\pi^2 c_w^2 s_w^2}\left( c_w^4 b^{(2)}_F +s_w^4 b_F^{(1)} Y_L^2 d_F^{(2)}\right) \log {{M}_L^2({\cal S}) \over \mu^2}\right] Z_{\mu\nu} Z^{\mu\nu} \ ,
\ee
where $b_F^{(2)}=2/3$, $b_F^{(1)}=4/3$, $N_c$ is the dimensionality of the $SU(3)_c$ representation $L$ belongs to,
and $d_F^{(2)}=2$ is the dimensionality of the $SU(2)$ fundamental representation. In addition, 
$c_w$ and $s_w$ are the cosine and sine of the Weinberg angle.
Then we compute
\bea
c_2^{L}&=&\frac{\alpha_{em}}{3\pi} \frac{N_c}{c_w^2 s_w^2} \left(c_w^4 + 2 s_w^4 Y_L^2 d_F^{(2)}\right) \frac{m_S}{m_L} y_L  \ , \\
\Gamma^{(f)}(S\to ZZ) &=&  {\cal P}\left({m_Z^2\over m_S^2}\right) \frac1{64\pi}\,(c_2^L)^2 m_S  \label{eq:sZZf}\\
 \! \! &=& {\cal P}\left({m_Z^2\over m_S^2}\right)\, \frac{\alpha_{em}^2}{576\pi^3} \frac{N_c^2}{c_w^4 s_w^4}\left(c_w^4 + 2 s_w^4 Y_L^2 d_F^{(2)}\right)^2  \frac{m_S^3}{m_L^2} y_L^2 \  ,
 \eea
 where ${\cal P}(x)=\sqrt{1-4x}\, (1-4x+6x^2)$ is a factor correcting for the massive final states in the decay width. Notice the additional
 difference between Eq.~(\ref{eq:sZZg}) and Eq.~(\ref{eq:sZZf}) due to a  color factor of 8 since there are eight gluons in the 
 final states for $S\to gg$.
 
It is worth commenting that, since $SU(2)_L$ is broken and gauge invariance does not forbid a mass term for the $Z$ boson, one might expect a contribution to $c_1$ be
generated at the one-loop level. However, recall that 
 vector-like fermions do not give one-loop corrections to the $Z$ boson mass term, and thus make no contributions to the $m_Z^2 Z_\mu Z^\mu$ operator which would have given a contribution to $c_1$ after applying the Higgs low-energy theorem.
This argument suggests that any contribution to $c_1$ at one-loop level would come from applying the Higgs low-energy theorem to operators with four-derivatives such as $(\Box\, Z_\mu)^2$, which upon using the equation of motion is suppressed by $(m_Z/m_L)^4$ and can be safely neglected for a heavy $m_L$. We explicitly computed the
fermion triangle loop diagram in a large mass expansion, $m_L\to \infty$, and verified that the first contribution to $c_1$ indeed starts at $(m_Z/m_L)^4$.

\subsection{Gauge Boson Loop-induced $S\to ZZ$}
The last possibility we consider is when the messenger sector includes a new set of heavy electroweak gauge bosons $(W^{\prime}, Z^\prime)$. In the SM the $W$ contribution to the loop-induced decay of the Higgs into $\gamma\gamma$ dominates over the one from the top-quark loop due to a large beta function coefficient ``7'' multiplying the $W$ loop result~\cite{Ellis:1975ap}. One may expect a similar situation for the $W^\prime$ contribution to the loop-induced decay into $ZZ$.
Given the existence of two sets of electroweak gauge bosons, the simplest model must contain two copies of gauged $SU(2)$.
Schematically, the symmetry breaking pattern is a two-step process:
\be
SU(2)_1 \times SU(2)_2 \times U(1)_Y \rightarrow SU(2)_{L}
\times U(1)_Y \rightarrow U(1)_{em} \ ,
\ee
where the two $SU(2)$'s
are broken down to the vectorial subgroup, identified with $SU(2)_L$, at a high scale $f_1$ using a linear sigma model. Subsequently $SU(2)_L\times U(1)_Y$ is broken down to $U(1)_{em}$ at a low scale $f_2=v$ 
following the Higgsless model~\cite{Csaki:2003dt}.

In the Appendix A we explicitly construct a toy
model whose gauge sector is the same as the three-site Higgsless model~\cite{Casalbuoni:1985kq,Chivukula:2006cg,Foadi:2003xa}, although we are interested in a different corner of parameter space, $\epsilon \equiv (f_2/f_1)^2 \ll 1$. For example, if $f_1 \approx 1$ TeV and $f_2=v\approx 246$ GeV, we have $\epsilon \approx 0.06 \ll 1$. In this case the $W^\prime$ and $Z^\prime$ can be as light as several hundreds GeV for weakly
coupled theories. We computed the $W^\prime$ contribution to the $ZZ$ self-energy. At leading order in $\epsilon$,
\be
-\frac14\left[1- \frac{e^2}{16\pi^2 c_w^2 s_w^2}(7 c_w^4) \log {M_{W^\prime}^2({\cal S}) \over \mu^2}\right] Z_{\mu\nu} Z^{\mu\nu} \ ,
\ee
where we see the same large coefficient as in the scalar coupling to two photons. In the set up we have in the Appendix A,  
$M_{W^\prime}^2({\cal S}) = \frac12 (g_1^2+g_2^2)(f_1 + {\cal S})^2$ at leading order in $\epsilon$,
which leads to 
\be
\label{eq:c2wp}
c_2^{W^\prime} = \frac{7\alpha_{em}}{2\pi} \frac{c_w^2}{s_w^2} \frac{m_S}{f_1}  \ .
\ee
Using Eq.~(\ref{eq:sZZf}) we arrive at
\be
\label{eq:sZZW}
\Gamma^{(W^\prime)}(S\to ZZ) ={\cal P}\left({m_Z^2\over m_S^2}\right) \frac{49 \alpha_{em}^2}{256\pi^3} \frac{c_w^4}{s_w^4} \frac{m_S^3}{f_1^2}
\ee
There is also a contribution to $c_1$ induced at one-loop by the $W^\prime$ boson that is suppressed by $\epsilon$, which we ignore.

\phantom{some space here}

Given Eqs.~(\ref{eq:sZZg}), (\ref{eq:sZZf}), and (\ref{eq:sZZW}), we can now compare $\Gamma(S\to gg)$ with $\Gamma(S\to ZZ)$ and
see that the decay width into $ZZ$ can easily be comparable to the decay width into two gluons. 
This is especially the case for  the $W^\prime$ loop 
 due to the large coefficient in Eq.~(\ref{eq:c2wp}):
\be
\frac{\Gamma^{(W^\prime)}(S\to ZZ)}{\Gamma(S\to gg)} \sim 0.75 \times {\cal P}\left({m_Z^2\over m_S^2}\right)\, \frac{m_Q^2}{f_1^2 y_Q^2} \ .
\ee
Even in the case of purely fermionic contribution in $S\to ZZ$, assuming the fermion $(L^c, L)$ carries no hypercharge, the ratio of the two partial widths is 
\be
\frac{\Gamma^{(f)}(S\to ZZ)}{\Gamma(S\to gg)} \sim 0.01 N_c^2 \times
{\cal P}\left({m_Z^2\over m_S^2}\right)\, \frac{m_Q^2 y_L^2}{m_L^2 y_Q^2} \ ,
\ee
which could still be ${\cal O}(0.1)$ if the multiplicity factor $N_c \agt 3$. This could be achieved if the fermion $(L^c, L)$ is also in the fundamental representation of $SU(3)_c$, resulting in $N_c=3$. In the end, we have demonstrated that 
new physics scenarios giving rise to loop-induced decays of $S$ into gauge bosons could easily give a significant branching ratio
into $ZZ$ bosons. 

\section{Observables and Simulations}

In this section we discuss two observables which could be useful in disentangling whether the scalar coupling to $Z$ bosons is as predicted by the Higgs mechanism or induced by new physics at the loop-level.

\subsection{The Line Shape}

\begin{figure}
\includegraphics[scale=0.5, angle=0]{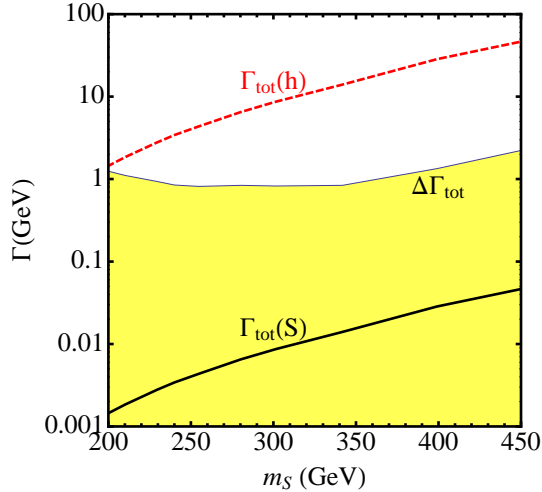}  
\caption{\label{fig2}{\em The dashed line is the total decay width for a SM Higgs boson and the solid line is that of a scalar $S$ whose
width is three orders of magnitude smaller. The yellow (shaded) region is the detector resolution.
}}
\end{figure}

Among the three possible on-shell couplings of $S$ with $Z$ bosons, only $c_1$ could be present at the tree-level with an order unity coupling 
when $S$ plays the role of the
Higgs boson in the Higgs mechanism, while both $c_2$ and $c_3$ are non-zero only at one-loop level. This observation suggests that the total
width of a scalar decaying through $c_2$ and $c_3$ must be much smaller than that of a scalar decaying through $c_1$, in order for the decay channel to be
observable in the early LHC running.
Using the SM Higgs as an example, the partial decay width $\Gamma(h\to VV)$ is
\be
\delta_V {\cal P}\left({m_V^2\over m_h^2}\right) \frac{G_Fm_h^3}{16 \sqrt{2}\pi}    \ , \qquad V= W, Z \ ,
\ee
where $G_F$ is the Fermi constant and $\delta_W=2\delta_Z=2$.  Comparing with the partial decay width of a $W^\prime$-loop induced decay, we see
\be
\frac{\Gamma^{(W^\prime)}(S\to ZZ)}{\Gamma(h\to ZZ)} \sim 10^{-3} \ ,
\ee
for a $W^\prime$ mass as light as 500 GeV.
As emphasized previously, in order for the event $gg\to S\to ZZ$ to be observable at the LHC with say 30 fb$^{-1}$ luminosity, the branching ratio ${\rm Br}(S\to ZZ)$ should be sizable and
comparable to that  of a SM Higgs into $ZZ$. It then follows that
\be
\frac{\Gamma_{\rm tot}(S)}{\Gamma_{\rm tot}(h)} = \frac{\Gamma(S\to ZZ)}{{\rm Br}(S\to ZZ)} \times \frac{{\rm Br}(h\to ZZ)}{\Gamma(h\to ZZ)}
 \sim 10^{-3} \ .
 \ee
In other words, we would observe an extremely narrow peak in the $ZZ$ invariant mass spectrum if the scalar $S$ only decays via $c_2$ and $c_3$. 
In fact, the peak is so narrow that the width is completely below the detector 
resolution. In this study we use a $2~\rm{GeV}$ bin size 
which is comparable to the energy resolution of the detector.
On the other hand, a scalar participating in the EWSB like the Higgs boson would have a width above the detector resolution at
the LHC except when its mass is below 200 GeV. Therefore, the Breit-Wigner line shape of a scalar
resonance in the $ZZ$ invariant mass spectrum is a strong indicator on the Higgs nature (or the lack thereof) of the scalar.

\begin{figure}
\includegraphics[scale=0.5, angle=0]{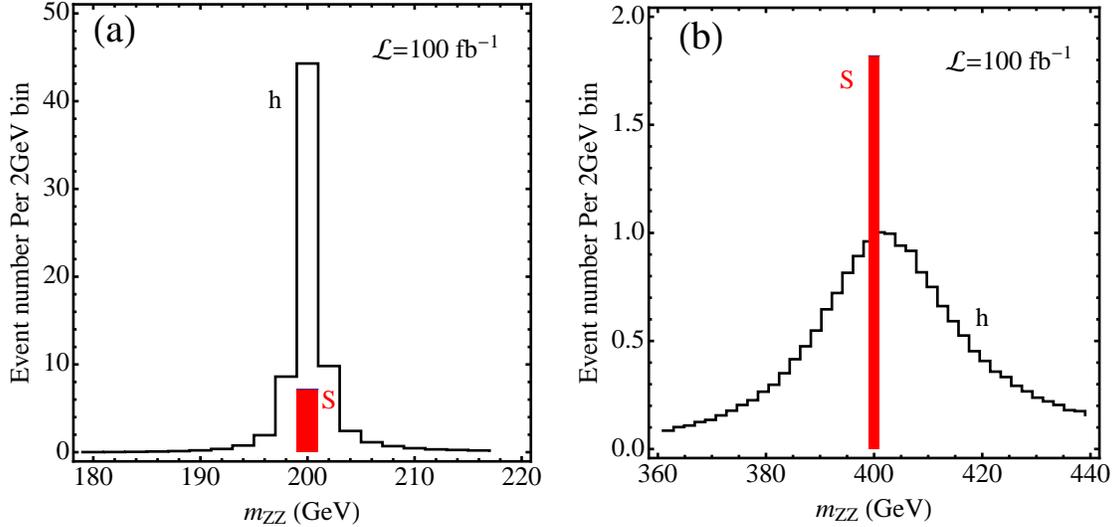}  
\caption{\label{fig3}{\em The $ZZ$ invariant mass distribution for a SM Higgs boson and a scalar $S$ decaying through loop-induced effects, using a 2 GeV bin size. The
narrow width of $S$ is below the detector resolution, resulting in a concentration of all events in just one bin. 
Note that for a sufficiently small bin size
one would resolve the peak, albeit with a form which is dominated by
the detector resolution (a Gaussian, if the usual assumptions of detector
smearing are made). 
In the plot we assume the event rate of $gg\to S\to ZZ\to 4\ell$ is
only 10\% of rate for the SM Higgs boson.
}}
\end{figure}

In Fig.~\ref{fig2} we show the total decay widths of a SM Higgs boson and a scalar $S$ decaying through loop-induced operators, and compare
 with the detector resolution at the LHC using the following lepton energy smearing:
 \be
 \frac{\delta E}{E} =  \frac{a}{\sqrt{E/{\rm GeV}}} \oplus b\ ,
 \ee 
where $a=13.4$\%, $b=2$\%, and $\oplus$ denotes a sum in quadrature~\cite{Abazov:2007ev}.
We see while the width of the SM Higgs could be resolved above a 200 GeV mass, the small width of the $S$ is
completely buried in the detector resolution. The narrow width of $S$ implies, in the invariant mass distribution of the two $Z$ bosons, all the events would be
concentrated in just one bin, resulting in a spectacular resonance peak even if the event rate is smaller than that of a SM Higgs. 
However, for a sufficiently small bin size,
one would resolve the peak, albeit with a form which is dominated by
the detector resolution (a Gaussian, if the usual assumptions of detector
smearing are made).  
In Fig.~\ref{fig3} we
simulate the $ZZ$ invariant mass distribution for a SM Higgs and the $S$ scalar using a 2 GeV bin size. To be conservative, in the plot
we assume the event rate $B\sigma(gg\to S \to ZZ \to 4\ell)$ is only 10\% of the SM Higgs. It is then clear that the total width measurement would allow for a distinction between the Higgs and a scalar $S$ which decays only at the loop-level, except when the Higgs has a mass below 200 GeV and its width is comparable to the detector resolution.

\subsection{Angular Distributions in $\phi$}
\label{sect:angphi}

\begin{figure}
\includegraphics[scale=0.65, angle=0]{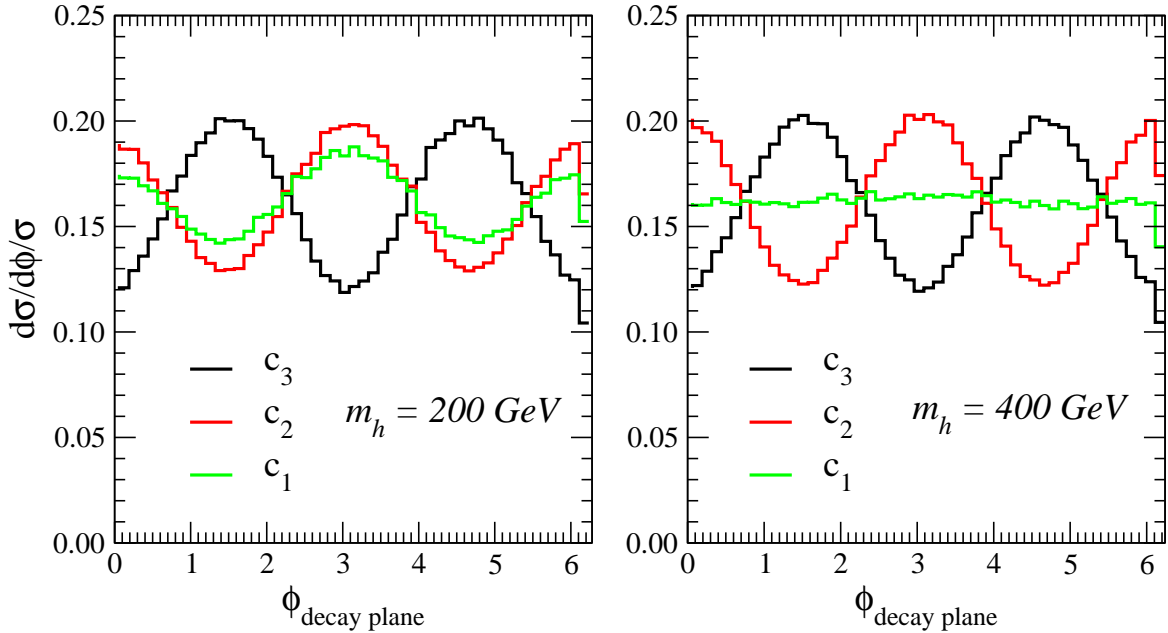}  
\caption{\label{fig4}{\em  The normalized azimuthal angular distributions for 200 and 400 GeV scalar masses, turning on one operator at a time.}}
\end{figure}

In the following we consider the dependence on the azimuthal angle between the two $Z$ decay planes in the normalized differential rate in Eq.~(\ref{eq:genphi}),
by turning on one operator at a time. By considering the normalized rate, the dependence on the magnitude of the coefficients $c_i$ in Eq.~(\ref{eq:effeL}) drops
out and the angular dependence is largely determined by kinematics. It is worth mentioning that in Eq.~(\ref{eq:genphi}) the $\cos(\phi + \delta)$ term is highly 
suppressed due to the approximate symmetry $g_L^2 \approx  g_R^2$ in the leptonic decays, so only the $\cos(2 \phi + 2 \delta)$ term and the constant term will 
contribute. As can be seen from Eq.~(\ref{eq:c1neq}), for a Higgs-like scalar, $c_1\neq 0$, the constant term becomes more dominant as the mass 
gets larger. On the other hand, for $c_2\neq 0$ the $\cos 2\phi$ terms is more important for a heavy scalar.

In Fig.~\ref{fig4} we simulate the azimuthal angular dependence in the normalized decay distribution for two different scalar masses: 200 and 400 GeV. 
To facilitate future experimental analysis, we provide in Appendix B Lorentz-invariant expressions for various angular variables in the decay into
$ZZ\to 4\ell$, including the azimuthal angle 
$\phi$.
We select
the events by requiring the following cuts on the lepton transverse momentum $p_T$ and rapidity $\eta$:
\be
p_T \ge 15 \ \ {\rm GeV}  \qquad {\rm and} \qquad  |\eta| \le 2.4 \ .
\ee
In the simulation we assume the SM background coming from $q\bar{q}\to ZZ\to 4\ell$ has been reduced in the data sample using existing procedures for
searching for a SM Higgs boson~\cite{Ball:2007zza}. The angular dependence of the SM background and its interplay in the Higgs search was studied in 
Ref.~\cite{Matsuura:1991pj}.
 In the case of loop-induced decays, one can take advantage of the extremely narrow width and
impose stringent cuts on the $ZZ$ invariant mass to reduce the backgrounds. Therefore we do not include backgrounds in the plot.

From Fig.~\ref{fig4} we see that the CP-odd case can be distinguished from the CP-even case in the angular distribution, which has been discussed in 
Refs.~\cite{Barger:1993wt, Godbole:2007cn}. The comparison of tree-level versus loop-induced operators in the CP-even case, however, does not seem to exist in the literature, to the best of our knowledge. We see that, even though $c_1\neq 0$ and $c_2\neq 0$ have the same phase in the angular distribution, the magnitudes are different even for a low mass of 200 GeV. Recall that this is also the mass range where the width measurement could be biased by the detector resolution.
So one could use the angular distribution as an extra handle to distinguish a Higgs boson from a non-Higgs-like scalar.

\section{Conclusion}

In this work we considered the most general on-shell couplings of a scalar with two $Z$ bosons. In the SM the decay of the Higgs boson into $ZZ$ final states
is the gold-plated mode for discovery due to excellent energy resolution for charged leptons. However, in order to verify the Higgs mechanism as the origin of
mass for the electroweak gauge bosons, it is necessary to measure the coupling between the scalar and the gauge bosons. By using an operator analysis for
the most general couplings, we point out that dimension-five operators responsible for the anomalous Higgs couplings are generated only at the loop-level,
while the Higgs mechanism would lead to a dimension-three operator at the tree-level.

Using the method of helicity amplitudes, we computed the differential decay distribution of a scalar decaying into $ZZ\to 4\ell$. Our formulas are simpler than
and agree with previous calculations. Furthermore, our results make clear the advantage of using the azimuthal angle between the two decay planes of the
$Z$ bosons in discerning effects between CP-odd and CP-even operators.

If the scalar is produced in the gluon fusion channel, which gives the largest production cross-section for the Higgs boson at the LHC, a decay channel into 
two gluons must also exist. Then in order for the event $gg\to S \to ZZ$ to be observable at the LHC in the early running, the partial decay width should be comparable to the partial width into two gluons. We investigated new physics scenarios giving rise to such a possibility by considering fermion-loop and $W^\prime$-loop induced couplings of
a scalar with $ZZ$ bosons.

 One important implication of loop-induced operators is that the total width of a non-Higgs-like scalar, if its decay were discovered at the LHC early on, should be order-of-magnitude smaller than that of a Higgs-like scalar, which decays through tree-level processes. Again this is a corollary of requiring a sizable branching ratio into
 $ZZ$ final states from loop-induced effects. Therefore measurements of the total width of a scalar resonance in final states with two $Z$ bosons is a strong indicator
 on the Higgs nature of the resonance, except when the scalar mass is below 200 GeV and the SM Higgs width is comparable to detector resolution. In this regard, azimuthal
 angular distribution could provide an extra handle in determining not only the CP property of the scalar but also whether the decay is loop-induced. Only when 
 the scalar coupling with the $Z$ bosons is verified to be the one as predicted by the Higgs mechanism, can one gain confidence in the Higgs mechanism as the
 origin of electroweak symmetry breaking as well as the discovery of a Higgs boson.

\section{Acknowledgements}
This work was supported in part by the U.S. Department of Energy under contract DE-AC02-06CH11357 (Argonne), and  by the World Premier International Research Center Initiative (WPI initiative) by MEXT, Japan.
Q.~H.~C. is supported in part by the Argonne National Laboratory and 
University of Chicago Joint Theory Institute (JTI) Grant 03921-07-137, 
and by the U.S.~Department of Energy under Grants No.~DE-AC02-06CH11357 
and No.~DE-FG02-90ER40560. 
J.~S. was also supported by the Grant-in-Aid for scientific research (Young Scientists (B) 21740169) from JSPS.
I.~L. acknowledges the hospitality of IPMU at the University of Tokyo while part of this work was performed.

\section*{Appendix A: $ZZ$ Self Energy From the Heavy Gauge Boson Loop}

In this Appendix, we compute the one-loop corrections to the $Z$ self energies which are needed to construct the $SZZ$ effective coupling  using the low-energy Higgs theorem.  For concreteness, we consider a simple gauge extension of the Standard Model (SM) which is based on the gauge group $SU(2)_1 \times SU(2)_2 \times U(1)_Y$. We use two link fields $\Sigma_1$ and $\Sigma_2$, 
\bea
\Sigma_1 &=& \frac{S}{f_1} e^{i\pi_1^a \sigma^a/f_1} \ ,  \qquad \langle S \rangle =f_1\ , \\
\Sigma_2 &=& e^{i\pi_2^a \sigma^a/f_2}  \  ,
\eea
which transform as bidoublets under $SU(2)_1\times SU(2)_2$ and  $SU(2)_1\times U(1)_Y$, respectively. The remaining unbroken gauge
group is identified with $U(1)_{em}$, whose
generator  is
$Q = T_3^{(1)} + {T}_3^{(2)} + Y/2$.  Notice that the gauge sector of this model is identical to the so-called three-site Higgsless model studied in Refs.~\cite{Casalbuoni:1985kq,Chivukula:2006cg,Foadi:2003xa}, except that we are allowing for a scalar degree of freedom in the radial 
excitation of $\Sigma_1$. More importantly, we are interested in the limit $\epsilon \equiv (f_2/f_1)^2 \ll 1$, which is also different from the three-site
Higgsless model.

The covariant derivatives are written as
\bea
D_\mu \Sigma_1& = &\partial_\mu \Sigma_1 - ig_1 \frac{\sigma^a}2 W_{1\,\mu}^{a} \Sigma_1 + i  \Sigma_1
 g_2 \frac{\sigma^a}2 W_{2\,\mu}^{a} \ ,\\
D_\mu \Sigma_2& = &\partial_\mu \Sigma_2 - i g_2 \frac{\sigma^a}2 W_{2\,\mu}^{a} \Sigma_2 + i  \Sigma_2
g^\prime \frac{\sigma^3}2 B_\mu ,
\eea
where $W_{i\,\mu}^{a}$ and $g_i$ are the gauge fields and coupling strengths belonging to $SU(2)_i$, $i=1,2$, respectively. 
Similarly $B_\mu$ and $g^\prime$ correspond to the gauge field and coupling of the $U(1)_Y$.
 The gauge bosons in the model obtain masses through the kinetic terms
 \be
 \frac{f_1^2}2 (D_\mu \Sigma_1)^\dagger (D^\mu \Sigma_1) +  \frac{f_2^2}2 (D_\mu \Sigma_2)^\dagger (D^\mu \Sigma_2),
 \ee 
 which lead to the the mass matrix for neutral gauge bosons $(W_1^3, W_2^3, B)$:
\be
\frac{1}{2} \left (
\begin{matrix}
 g_1^2 f_1^2 & -g_1 g_2 f_1^2 & 0 \cr
-g_1 g_2 f_1^2 & g_2^2 (f_2^2 + f_1^2) & -g' g_2 f_2^2 \cr
0 & - g' g_2 f_2^2 & g'^2 f_2^2 \cr 
\end{matrix} \right ) \ .
\ee
This matrix can be
diagonalized by means of an orthogonal matrix which we shall call {\bf R}:
\be
\begin{pmatrix} 
W_{1\mu}^3 \cr W_{2\mu}^3 \cr B_\mu \cr 
\end{pmatrix} = {\rm \bf R^\dagger }
\begin{pmatrix}  A_\mu \cr  Z_\mu \cr  Z^\prime_\mu \cr 
\end{pmatrix} \ ,
\ee
where the mass eigenstates are denoted by $A$, $Z$, and $Z^\prime$. The
eigenstate $A$ is massless and identified as the photon. The couplings
of our theory are related to the electric charge
by
\be
g_1 = \frac{e}{\cos \phi \sin \tw} \, , \quad
g_2 = \frac{e}{\sin \phi \sin \tw} \, , \quad
g' = \frac{e}{\cos \tw}
\ee
where $\tw$ is the weak mixing angle (in the limit $\epsilon \to 0$) and $\phi$ is an additional mixing
angle.  The other two eigenmasses are
\bea
m_Z^2 &=& \frac12 f_2^2 (g^2+g^{\prime 2}) \left[1 - \epsilon f_1^2 \frac{g_2^4}{(g_1^2+g_2^2)} \right]\ ,\\
m_{Z^\prime}^2 &=& \frac12 f_1^2 (g_1^2+g_2^{2}) \left[1 - \epsilon f_1^2 \frac{g_2^4}{(g_1^2+g_2^2)} \right]  \ ,
\eea
where we have dropped ${\cal O}(\epsilon^2)$ terms and  
\be
\frac{1}{g^2} \equiv \frac1{g_1^2} + \frac1{g_2^2}\ .
\ee
Clearly, $Z$ is identified with the SM $Z$ boson while $Z^\prime$ is referred to as the heavy $Z$ boson.
For small $\epsilon$, the mixing matrix {\bf R} has the following approximate form:
\be
\footnotesize {\rm \bf R} = \left (
\begin{matrix}
\cos \phi \sin \tw &
\sin \phi \sin \tw & \cos \tw \cr \cos \phi \cos \tw
+ \epsilon \frac{\cos^3 \phi \sin^2 \phi}{\cos \tw} & \sin \phi \cos \tw
- \epsilon \frac{\sin \phi \cos^4 \phi}{\cos \tw} & - \sin \tw \cr
- \sin \phi + \epsilon \sin \phi \cos^4 \phi & \cos \phi
+ \epsilon \sin^2 \phi \cos^3 \phi  & - \epsilon \tan \tw \sin \phi \cos^3
\phi \cr 
\end{matrix}
\right )\ ,
\ee
from which it is simple to verify that tree-level couplings between the scalar $S$ and the $Z$ boson start only at order $\epsilon^2$. In other words, 
in this model $c_1=0$ at tree-level if we only keep terms up to ${\cal O}(\epsilon)$. The charged gauge boson sector can be worked out in a similar way, where the light mass eigenstate is identified with the SM $W$ boson and the heavy eigenstate is denoted by  $W^{\prime}$. 
The couplings between $S$ and the $W^\prime$ and $Z^\prime$ 
bosons have the form as predicted by the Higgs mechanism:
\be
m_{V^\prime}^2\left(1+\frac{S}{f_1}\right)^2 V_\mu^\prime V^{\prime\,\mu} \ , \quad V=W, Z \, \ ,
\ee
which is valid at leading order in $\epsilon$.

We would like to compute the one-loop correction to the $Z$ self energy arising from loops of $W^\prime$ gauge bosons.  Unfortunately, much like the analogous corrections in the SM, the corrections to the two-point functions depend non-trivially on the particular $R_\xi$ gauge used to define the $W^\prime$ propagator~\cite{Degrassi:1992ff}.  However, by extracting $R_\xi$ gauge-dependent pieces from other one-loop corrections (i.e., vertex and box corrections) and summing these with those from the two-point function one can obtain an expression which is independent of the particular gauge chosen to do the calculation.  This method, which is known as the Pinch Technique (PT), has been applied to the SM to obtain gauge-independent expressions for the gauge boson self-energies~\cite{Degrassi:1992ue}.  More recently, though, it has been extended to models with extended gauge sectors such as the model considered here~\cite{Matsuzaki:2006wn,Dawson:2007yk,Sekhar Chivukula:2007ic}.  In this work, we will directly apply the results from the above references by taking the limit of our interest, $\epsilon=(f_2/f_1)^2 \ll 1$.  We refer interested readers to Ref.~\cite{Dawson:2007yk} for details and only make the following two comments. First, our results are obtained by
taking the so-called ``ideal localization" limit for the de-localized fermion introduced in Ref.~\cite{Chivukula:2006cg}.\footnote{In our model ideal localization is achieved by choosing the delocalization parameter $x_1$, which is defined
in Ref.~\cite{Chivukula:2006cg}, to be
$\sin^2\phi/(\sin^2\phi - \cos^2\phi)$.}
Such a limit has the advantage of reducing the tree-level $S$ parameter in the model. However, the main reason in our case is
to decouple the de-localized fermion from the $W^\prime$ boson, so as to remove the extra pinch contribution to the two-point function that is unnecessary for maintaining the gauge invariance. Second, even though we are allowing for a scalar degree of freedom in the radial excitation of
$\Sigma_1$, which is absent in the three-site Higgsless model, the computation in Ref.~\cite{Dawson:2007yk} still carries through because
$S$ has no couplings to the $Z$ boson in the order we are working. Therefore the $Z$ self energy in the non-linear sigma model (Higgsless model) is the same as in the (partially) linear sigma model we consider.

The one-loop expression for the $Z$ self-energy computed using the PT are then given by (up to ${\cal{O}}(\epsilon^2)$):
\bea
\Pi_{ZZ}^{(W^\prime)}(p^2)& =& \frac{\alpha_{em}}{4 \pi s_w^2 c_w^2}  \left[  - \frac{3}{2} \epsilon^2 \frac{m_{W^\prime}^4}{m_W^2} \cos^6\phi \sin^6\phi  \right.\nonumber \\ 
&& \left.\phantom{- \frac{3}{2} } +p^2 \left( 7c_w^4 - 14\,\epsilon\, \cos^2\phi \cos2\phi\, c_w^2\right) \right]  \log \frac{\Lambda^2}{m_{W^\prime}^2} \, ,
\eea
where $\Lambda$ is the cutoff of our effective theory.  Notice that 
formally $(m_{W^\prime}/m_W)^2 \sim 1/\epsilon$ so the longitudinal piece is considered $m_{W^\prime}^2\times {\cal O}(\epsilon)$, while
the leading term in the transverse component has a large coefficient ``7,'' the same as in the SM $W$ contribution to the photon self energy, which is to be expected.

\section*{Appendix B: A Lorentz-Invariant Construction of $\phi$}

In this Appendix we provide a Lorentz-invariant expression for the azimuthal angle $\phi$ between the two decay planes
of the $ZZ$ pair, so as to facilitate the analysis of angular distribution in $\phi$.
Let $p_1$ and $p_2$ be the momenta of the lepton pair coming from one  $Z$, and $p_3$ and $p_4$ be
the momenta of the lepton pair from the other $Z$. The parent momentum is $P=p_1+p_2+p_3+p_4$, which
satisfies the on-shell dispersion relation $P^2=M^2$.
We follow the  notation in Ref.~\cite{Keung:2008ve}, 
$p_1=\ell_1$, $p_2=\bar\ell_1$, $p_3=\ell_2$, $p_4=\bar\ell_2$. (See also Fig.~\ref{fig1}.)

In the rest frame of $P$, our azimuthal angle $\phi$ is given by
\be
\label{eq:phinoncov}
 {{\bf p}_1\times {\bf p_2} \over |{\bf p}_1|  |{\bf p}_2| \sin{\bar\theta}_{12}}
\cdot 
  {{\bf p}_3\times {\bf p_4} \over |{\bf p}_3|  |{\bf p}_4| \sin{\bar\theta}_{34}}
=-\cos\phi  \ ,
\ee
where the triple products in the numerator can be written in a Lorentz-invariant fashion:
\be
 ({\bf p}_1\times {\bf p_2})^i=\frac1M \epsilon^{\mu\nu i \rho} p_{1\mu} p_{2\nu} P_\rho \equiv  \frac1M \epsilon^{p_1p_2iP} \ , \qquad
  ({\bf p}_3\times {\bf p_4})^i=\frac1M\epsilon^{p_3p_4iP} \ .
\ee  
Note that we define $\epsilon^{1230}=1=\epsilon_{0123}=-\epsilon^{0123}$. Then it follows
\be
  ({\bf p}_1\times {\bf p_2})\cdot ({\bf p}_3\times {\bf p_4})
=-g_{\mu\nu}{\epsilon^{p_1p_2\mu P} \epsilon^{p_3p_4\nu P} \over M^2  }
= 
{1\over M^2} \left|
\begin{array}{ccc}
p_1\cdot p_3 & \ p_1\cdot p_4 \ & p_1\cdot P\\
p_2\cdot p_3 & \ p_2\cdot p_4 \ & p_2\cdot P\\
P  \cdot p_3 & \ P\cdot p_4   \ &   M^2 \end{array}\right|  \ .
\ee
To arrive at a covariant expression for Eq.~(\ref{eq:phinoncov}), we need to cast
the denominator in the covariant form as well:
\be 
|{\bf p}_1|=\frac1M \, p_1\cdot P \ ,\quad 
\cos{\bar\theta}_{12}=1-{m_{12}^2\over  2|{\bf p}_1| |{\bf p}_2|}  \ ,
\ee
where $m_{ij}^2 \equiv (p_i+p_j)^2$, 
and similarly for $|{\bf p}_2|$,$|{\bf p}_3|$, and $|{\bf p}_4|$. In the end
we have
\be
\cos\phi=-{M^2 \left|
\begin{array}{ccc}
p_1\cdot p_3 & \ p_1\cdot p_4 \ & p_1\cdot P\\
p_2\cdot p_3 & \ p_2\cdot p_4 \ & p_2\cdot P\\
P  \cdot p_3 & \ P\cdot p_4   \ &   M^2 \end{array}\right|
       \over 
(p_1\cdot P)(p_2\cdot P)(p_3\cdot P)(p_4\cdot P)
\sqrt{1-\left(1-{M^2m_{12}^2 \over 2 p_1\cdot P\, p_2\cdot P}\right)^2}
\sqrt{1-\left(1-{M^2m_{34}^2 \over 2 p_3\cdot P\, p_4\cdot P}\right)^2}
}\ .
\ee

On the other hand,
 $\sin\phi$ can be evaluated by the following
relation,
\be 
 \sin\phi=-\frac1M {\epsilon^{p_1p_2p_3p_4}  |{\bf p}_1+{\bf p}_2|
\over  |{\bf p}_1|\ |{\bf p}_2| \  |{\bf p}_3|\ |{\bf p}_4| 
\sin{\bar\theta}_{12}\sin{\bar\theta}_{34} } \ ,
\ee
where 
\be
\epsilon^{p_1p_2p_3p_4} 
=\epsilon_{\mu\nu\alpha\beta}\, p_1^\mu \, p_2^\nu\, p_3^\alpha\, p_4^\beta 
=\left|\begin{array}{cccc}
E_1 &   p_1^x & p_1^y & p_1^z \\
E_2 &   p_2^x & p_2^y & p_2^z \\
E_3 &   p_3^x & p_3^y & p_3^z \\
E_4 &   p_4^x & p_4^y & p_4^z \end{array}\right| 
= - M\, {\bf p}_3\times {\bf p_4} \cdot {\bf p}_1
\  . 
\ee
The covariant form is given by
\be
  \sin\phi=-\frac12 \frac{M^4 \lambda^{\frac12} \epsilon^{p_1p_2p_3p_4}}{
       (p_1\cdot P)(p_2\cdot P)(p_3\cdot P)(p_4\cdot P)
        \sqrt{1-\left(1-{M^2m_{12}^2 \over 2 p_1\cdot P\, p_2\cdot P}\right)^2}
         \sqrt{1-\left(1-{M^2m_{34}^2 \over 2 p_3\cdot P\, p_4\cdot P}\right)^2}} \ ,
 \ee
with $\lambda
\equiv1+m_{12}^4/M^4+m_{34}^4/M^4-2m_{12}^2/M^2
-2m_{34}^2/M^2-2m_{12}^2m_{34}^2/M^4$ .

We can also determine the polar angle of $p_1$ in the rest frame
of the 12 pair.
A simple Lorentz boost gives
\be       \bar E_1= \gamma E_1 (1+\beta\cos\theta) 
   = { \bar E_1 +\bar E_2\over m_{12}} {m_{12}\over2} (1+\beta\cos\theta_1)  \ ,
 \ee
which leads to 
\be
 \cos\theta_1={\bar E_1-\bar E_2\over |{\bf p}_1+{\bf p}_2|}
=  \frac2{M^2 \lambda^{1\over2}}(p_1\cdot P-p_2\cdot P )  \ .
\ee
For the polar angle of $p_3$ in the rest frame of the 34 pair, simply replace $p_1$ and $p_2$ by $p_3$ and $p_4$, respectively, in the above.

\end{document}